\documentclass[pss]{wiley2sp} 
\usepackage{amsmath}

\begin{document}

\title{Is there a common theme behind the correlated-electron superconductivity in organic charge-transfer
solids, cobaltates, spinels and fullerides?}

\titlerunning{Correlated-electron superconductivity}

\author{%
  Sumit Mazumdar\textsuperscript{\Ast,\textsf{\bfseries 1}},
  R. Torsten Clay\textsuperscript{\textsf{\bfseries 2}}}

\authorrunning{Mazumdar et al.}

\mail{e-mail
  \textsf{sumit@physics.arizona.edu}, Phone:
  +001-520-621-6803, Fax: +001-520-621-4721}

\institute{%
  \textsuperscript{1}\,Department of Physics, University of Arizona, Tucson, AZ 85721, USA\\
  \textsuperscript{2}\,Department of Physics and Astronomy and HPC$^{2}$ Center for Computational Sciences, Mississippi State University, Mississippi State, MS 39762, USA \\} 

\received{XXXX, revised XXXX
``````````````````````````, accepted XXXX} 
\published{XXXX} 


\abstract{ \abstcol{ We posit that there exist deep and fundamental
    relationships between the above seemingly very different
    materials.  The carrier concentration-dependences of the
    electronic behavior in the conducting organic charge-transfer
    solids and layered cobaltates are very similar. These dependences
    can be explained within a single theoretical model, - the extended
    Hubbard Hamiltonian with significant nearest neighbor Coulomb
    repulsion. Interestingly, superconductivity in the cobaltates seems
    to be restricted to bandfilling exactly or close to one-quarter,
    as in the}{organics. We show that dynamic Jahn-Teller effects and
    the resultant orbital ordering can lead to $\frac{1}{4}$-filled
    band descriptions for both superconducting spinels and fullerides,
    which show evidence for both strong electron-electron and
    electron-phonon interactions.  The orbital orderings in
    antiferromagnetic lattice-expanded bcc M$_3$C$_{60}$ and the
    superconductor are different in our model.  Strong correlations,
    quarter-filled band and lattice frustration are the common
    characteristics shared by these unusual superconductors. }}

\maketitle   

\section{Introduction.}
The field of superconductivity (SC) is facing a crisis: 25 years after
the discovery of high T$_c$ SC, none of the proposed scenarios has led
to a mechanism that scientists agree on. Although there is now general
agreement that there exist many correlated-electron superconductors,
there is no understanding how any one scenario, proposed for one class
of materials, could be extended to others. In the context of organic
charge-transfer solids (CTS) alone, the natures of the insulating
states proximate to SC can be quite different, including spin-density
wave, antiferromagnetic, charge-ordered or even a so-called valence
bond solid. It is unlikely that the BCS theory, designed to explain
the metal-to-superconductor transition, applies to any of these
insulator-superconducting transitions. It is equally unlikely,
however, that the mechanism of SC in structurally similar CTS, with
closely related molecular components, are different for different
initial insulating states, as has sometimes been proposed.  We believe
that determination of the characteristics shared by many seemingly
different {\it classes} of superconducting materials will give the
{\it framework} within which the theory of exotic SC should be
constructed. Our goal here is to demonstrate that such a common
framework might indeed exist for the CTS, inorganic layered cobaltates
and spinels, and fullerides.

In the next section we examine the counterintuitive carrier
density-dependent electronic behavior in the layered cobaltates.  We
point out that this behavior is very similar to that previously noted
in the CTS, and show that both cobaltates and CTS can be understood
within the same theoretical model, within which the bandfilling is a
very important implicit parameter. It is then noteworthy that in both
families SC appears to be limited to the $\frac{1}{4}$-filled
band. Other features shared by the superconducting cobaltates and
organics are strong Coulomb correlations and lattice frustration.  We
show that spinels and fullerides share the features common to the CTS
and cobaltates.  We suggest that for the strongly correlated
frustrated $\frac{1}{4}$-filled band the Schafroth mechanism of SC
\cite{Schafroth55a} becomes operative.
\begin{figure}[tb]%
\centerline{\includegraphics*[width=7.3cm,,height=5.2cm]{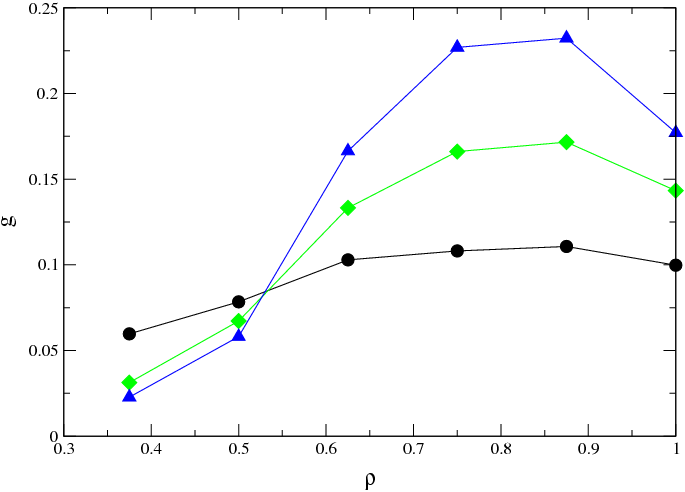}}
\caption{(Color online) Exact $g(\rho)$ versus $\rho$ for $U=10$,
and $V=0$ (circles), 2 (diamonds) and 3 (triangles), for periodic ring of 16 sites. Lines are guides to the
eye.}
\label{g-1d}
\end{figure}

\section{Layered cobaltates and CTS - carrier density dependence.}

We are interested in M$_x$CoO$_2$ (M = Li, Na, K) \cite{Wang03a}, the
incommensurate ``misfits'' [Bi$_2$A$_2$O$_4$][CoO$_2$]$_m$ (A = Ca,
Sr, Ba) \cite{Brouet07a} as well as the hydrated superconductor
Na$_x$CoO$_2$, $y$H$_2$O \cite{Takada03a}.  These materials consist of
CoO$_2$ layers, with the Co-ions forming a triangular lattice,
separated by layers of M$^+$ ions. The Co$^{3+}$ and Co$^{4+}$ ions
are in their low-spin states because of large crystal field splitting
\cite{Hasan04a}. Trigonal distortion splits the $t_{2g}$ orbitals into
low lying fully occupied doubly degenerate $e_g^\prime$ orbitals and
higher energy $a_{1g}$ orbitals that are occupied by one (in
Co$^{4+}$) or two (in Co$^{3+}$) electrons \cite{Mizokawa04a}.  Only
the $a_{1g}$ orbitals are relevant for transport and
thermodynamics. Charge carriers are $S=\frac{1}{2}$ holes on the
Co$^{4+}$ \cite{Hasan04a}, and hence the carrier density $\rho=1-x$ in
M$_x$CoO$_2$.

The onsite repulsion Hubbard $U$ is large in cobaltates
\cite{Wang03a}.  Simplistically, large $\rho<1$ is close to the
$\rho=1$ Mott-Hubbard limit and should behave as strongly correlated;
small $\rho$ should behave as weakly correlated as it is close to the
$\rho=0$ band semiconductor limit (all Co-ions trivalent.) Experiments
probing magnetic susceptibility and thermoelectric power have,
however, demonstrated that the {\it actual $\rho$-dependence is
  exactly the opposite}; in the experimentally accessible range
$\rho=0.2-0.8$, small (large) $\rho$ is strongly (weakly) correlated
\cite{Wang03a,Brouet07a}. The basic observation is the same for all
M$_x$CoO$_2$ and the misfits, indicating that $\rho$-dependence is
intrinsic to the CoO$_2$ layer.

A hint to the understanding of cobaltates comes from prior observation
on quasi-one-dimensional (quasi-1D) conducting CTS \cite{Mazumdar}, in which
$\rho$ ranges from 0.5 to 1. CTS with
$\rho$ equal to or close to 0.5 exhibit magnetic susceptibility
enhanced relative to the Pauli susceptibility, and lattice instability
with periodicity 4k$_F$ (where k$_F$ is the Fermi wavevector within
one-electron theory), both accepted as signatures of strong
correlations. In contrast, $\rho$ between 0.66 and 0.8 show weakly
correlated behavior, viz., unenhanced susceptibility and the usual
2k$_F$ Peierls instability (see Tables in reference \cite{Mazumdar}).
The behavior in the CTS and the cobaltates are then very similar.

\section{Theory of carrier density-dependent electronic behavior.}

We present here a theory of the $\rho$-dependent electronic behavior,
in 1D \cite{Mazumdar} and the 2D triangular lattice
\cite{Li11a}. Consider the extended Hubbard Hamiltonian,
\begin{equation}
H=-\sum_{\langle ij \rangle \sigma}t_{ij}c^\dagger_{i\sigma}
c_{j\sigma} + U\sum_{i} n_{i,\uparrow}n_{i,\downarrow} 
+ V \sum_{\langle ij \rangle} n_{i} n_{j}
\label{ham}
\end{equation}
where $c^\dagger_{i\sigma}$
creates an electron or a hole with spin $\sigma$ ($\uparrow$ or $\downarrow$) on site $i$, $\langle ... \rangle$
implies
nearest neighbors (NN), and all other terms have their usual meanings.
Individual molecules (Co-ions) are the sites in the CTS (cobaltates).
\begin{figure}[t]%
\centerline{\resizebox{8cm}{!}{\includegraphics{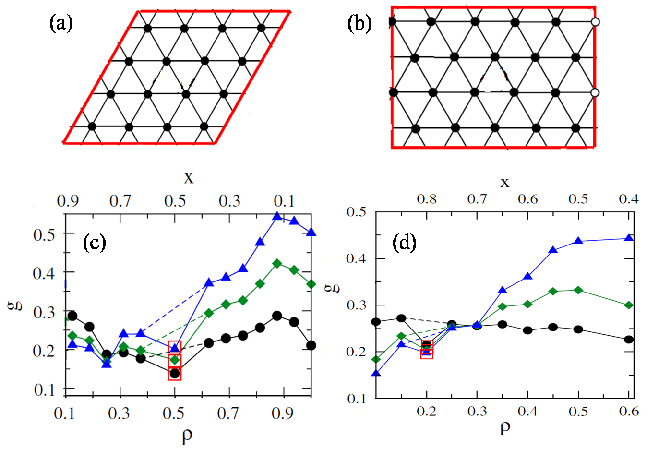}}}
\caption{(Color online) (a) and (b) $N_s=16$ and 20-site clusters
  investigated numerically.  (c) and (d) Exact $g(\rho)$ versus
  $\rho$. The circles, diamonds and triangles correspond $V=0$, 2 and 3,
  respectively. The boxes correspond to data points with
  total spin $S>S_{\rm{min}} = 0(\frac{1}{2})$ for even (odd) numbers
  of particles.  }
\end{figure}

\begin{figure*}[htb]%
\sidecaption
\resizebox{0.69\textwidth}{!}{\includegraphics{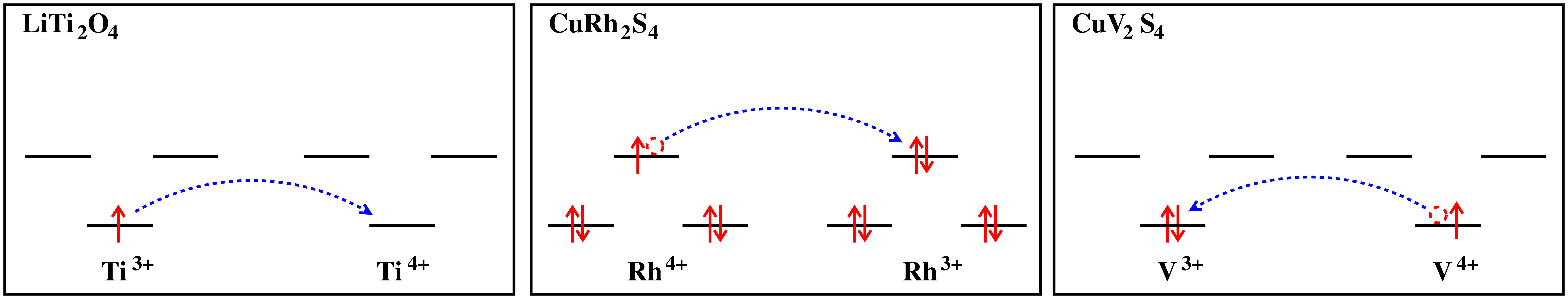}}
\caption{(Color online) The effective $\frac{1}{4}$-filled band natures of the
  superconducting spinels, following the splitting of the $t_{2g}$
  orbitals (see text). The blue dotted arrows denote electron or hole
  motion.}
\label{twocolumnfigure}
\end{figure*}
Strongly correlated behavior, at any $\rho$, requires that the ground
state wavefunction has relatively small contribution from
configurations with double occupancies.  We therefore calculate
numerically the normalized probability of double occupancy in the
ground state
\begin{equation}
g(\rho)=\frac{\langle n_{i,\uparrow}n_{i,\downarrow}\rangle}{\langle n_{i,\uparrow}\rangle \langle n_{i,\downarrow}\rangle}
\end{equation}
$g(\rho)$ is clearly weakly $\rho$-dependent for $V=0$. For $V>0$ and
$\rho=0.5$, electrons are prevented from encountering one another by
$V$, thus reducing $g(\rho)$ more than $U$ alone. As $\rho$ is
increased from 0.5, more and more intersite repulsions are generated
with additions of electrons, and configurations with a few double
occupancies begin to compete with those with only singly occupied
sites.  Thus $g(\rho)$ increases steeply with increase in $\rho$ for
significant $V/U$.  As $\rho$ increases beyond an intermediate range,
the Mott-Hubbard limit is approached and $g(\rho)$ should decrease
again.

Exact numerical calculations of $g(\rho)$ verify this conjecture. Our
earlier 1D calculations were for relatively small numbers of electrons
$N_e$ with varying number of sites $N_s$ \cite{Mazumdar}.
Significantly larger computer capabilities now allow calculations for
different $\rho$ with fixed $N_s=16$. In Fig.~1 we have shown the
results of our calculations for $U/|t|=10$ and $V/|t|$=0, 2 and
3. This value of $U$ is considered realistic for CTS, and there is
evidence for $V/|t|$ as large as 3. The plots are similar for
$U/|t|=6-12$ and $V/|t|=1-3$. While CTS with $\rho<0.5$ do not exist,
we have included this smaller $\rho$ for comparison to 2D.  We
conclude from Fig.~1 that for realistic $V/U$, strong $\rho$-dependent
electronic behavior is warranted.  Experimentally \cite{Mazumdar},
{\it all} $\rho=0.5$ CTS (e.g., MEM(TCNQ)$_2$, Qn(TCNQ)$_2$,
(TMTTF)$_2$X, etc.)  exhibit strongly correlated behavior, while CTS
with $\rho=0.66-0.8$ (TTFBr$_{0.74}$, HMTTF- and HMTSF-TCNQ with
$\rho=0.75$, etc.) uniformly exhibit weakly correlated
behavior. Subsequent demonstrations of strongly correlated behavior in
2:1 BEDT-TTF \cite{Kanoda06a} or 1:2 Pd(dmit)$_2$ systems
\cite{Kato04a} further support the theory.

In Fig.~2 we show our results for the finite periodic 2D triangular
lattices with $N_s=16$ (Fig.~2(a)) and 20 (Fig.~2(b).) We have taken
$t>0$. Elsewhere we have argued that the $U/|t|$ and $V/|t|$ in the
cobaltates are similar to that in the CTS \cite{Li11a}. With the
exception of a few $N_e$, the ground state is always in the lowest
spin state (0($\frac{1}{2}$) for $N_e$=even(odd)) for $t>0$. As seen
in Figs.2(c) and (d) the behavior of $g(\rho)$ in 2D is similar to
that in 1D, and explains the much-discussed lack of symmetry about
$\rho=0.5$ in the cobaltates - strongly correlated behavior near
$\rho=\frac{1}{3}$ and weakly correlated behavior near
$\rho=\frac{2}{3}$ \cite{Wang03a}.

\section{Superconductivity in the CTS and layered cobaltates.}

We point out that SC is limited to the same narrow range of carrier
concentration in both families.  Given the strong role of $\rho$, this
cannot but be significant.  CTS superconductors are 2:1 cationic or
1:2 anionic materials \cite{Ishiguro}, with 100\% charge-transfer, and
consequently $\rho=0.5$. The lattice structures of the superconducting
CTS are not quasi-1D but anisotropic triangular, giving both
frustration and larger bandwidth.
  
In the hydrated superconducting Na-cobaltate, $x \simeq 0.35$, Careful
experiments have revealed however that $\rho$ is considerably smaller
than 0.65, as a portion of the water enters as H$_3$O$^+$.  A
significant number of investigators have determined that $\rho$ in the
superconductor is exactly or close to 0.5 \cite{Barnes05a}.  It is
also noteworthy that $x=0.5$ is unique, - Hall and Seebeck
coefficients change signs only at this $\rho$ at low temperature in
both Na$_x$CoO$_2$ and Li$_x$CoO$_2$ .

The limitation of SC to $\rho=0.5$ acquires additional significance
when taken together with: (a) absence of SC in the $\rho=1$ triangular
lattice Hubbard Hamiltonian \cite{Clay08a}, (b) the propensity at
$\rho=0.5$ to form the paired-electron crystal (PEC), in which
spin-singlet pairs are separated by pairs of vacant sites, in both 1D
and the moderately frustrated anisotropic triangular lattice
\cite{Clay11a}. We have proposed that with further increase in
frustration the PEC gives way to a superconducting paired-electron
liquid \cite{Clay11a}.  The proposed theory of SC is essentially the
same as that of Schafroth, within which electron-pairs form mobile
molecules (charged bosons) \cite{Schafroth55a}. It is, however, different from the
mean-field theory of charge-fluctuation mediated SC \cite{Merino01a}.  
In the rest of this
paper we show that there exist other systems which likely can be
understood within the same scenario.

\section{Superconducting spinels.}

Spinels are inorganic ternary compounds AB$_2$X$_4$, where the
X$^{2-}$ form a close-packed structure with the A (B) sites occupying
tetrahedral (octahedral) interstices. The B-cations are the active
sites, with small X-mediated B-B electron hoppings. The B sublattice
is tetrahedral, and hence frustrated.

LiTi$_2$O$_4$, CuRh$_2$S$_4$ and CuRh$_2$Se$_4$ are the only three
spinels that have been confirmed to be superconductors; reports of SC
in CuV$_2$S$_4$ and CoCo$_2$S$_4$ exist.  SC in LiTi$_2$O$_4$ was
discovered many years before the cuprates \cite{Johnston76a}, and has
been of interest ever since because of its high T$_c$/T$_F$ ($T_F$ is
Fermi temperature), which is closer to that in the cuprates than
conventional superconductors \cite{Wu94a}. The mechanism of SC remains
controversial: strong coupling BCS theory, the RVB approach
\cite{Anderson87a} and the bipolaron theory \cite{Alexandrov81a} of SC
have all been proposed.

Examination of the common valence of the B-cations in the
superconducting spinels, 3.5+, reveals remarkable similarities between
them and the CTS and cobaltates.  The Ti$^{3.5+}$ ions in
LiTi$_2$O$_4$ have one $d$-electron per two Ti.  Rh$^{3.5+}$ and
Co$^{3.5+}$ in their low-spin states possess one $d$-hole per two
metal ions. V$^{3+}$ and V$^{4+}$ have electron configurations 3d$^1$
and 3d$^2$, respectively. In all cases static band Jahn-Teller lattice
distortions will give $\frac{1}{4}$-filled electron (in LiTi$_2$O$_4$)
and hole (in CuRh$_2$S$_4$, CoCo$_2$S$_4$ and CuV$_2$S$_4$)
\begin{figure*}[htb]%
\sidecaption
\includegraphics*[width=.55\textwidth]{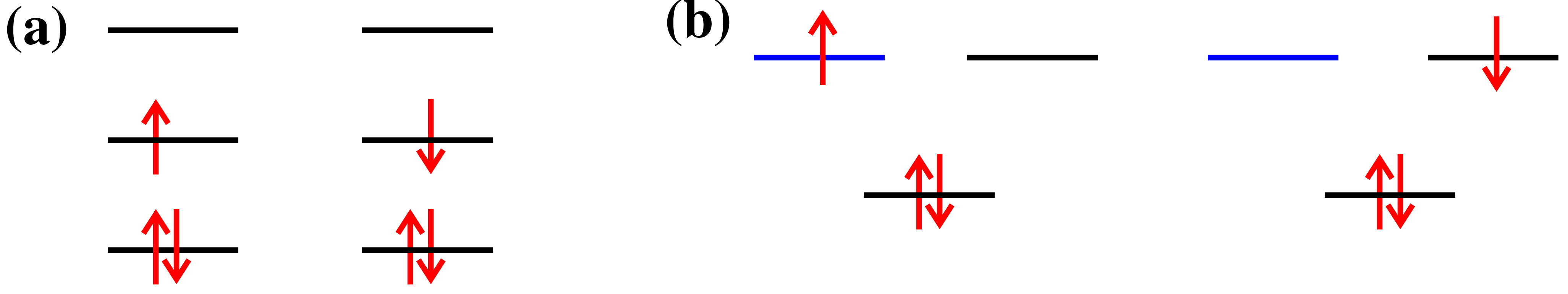}
\caption{(Color online) (a) The effective $\frac{1}{2}$-filled $t_{1u}$ band in antiferromagnetic C$_{60}^{3-}$,
following Jahn-Teller distortion. 
(b) The proposed doubly degenerate $\frac{1}{4}$-filled bands in the superconductor
(see text).
}
\end{figure*}
bands as indicated in Fig.~3.  Static lattice distortions should give
a 3D PEC, as indeed observed in CuIr$_2$S$_4$ \cite{Radaelli02a} and
LiRh$_2$O$_4$ \cite{Okamoto08a}. We propose that the superconducting
state is a 3D paired-electron liquid, with dynamical (as opposed to
static) orbital ordering.

\section{Superconducting fullerides.}

The C$_{60}^{3-}$ ions in superconducting M$_3$C$_{60}$ form fcc
lattices.  Theories of fullerides treat each C$_{60}$ unit as a site,
with electron hoppings between the triply degenerate $t_{1u}$
antibonding molecular orbitals.  Although many experiments suggest
onsite pairing mediated by $H_g$ Jahn-Teller phonons \cite{Varma91a},
the large T$_c$/T$_F$ suggests a non-BCS mechanism \cite{Wu94a}.
Antiferromagnetism in bcc Cs$_3$C$_{60}$ also indicates strong Coulomb
repulsion \cite{Iwasa03a,Capone09a}.  Nonzero spin gap to the lowest
energy high-spin state in the antiferromagnet has indicated that it is
the effective $\frac{1}{2}$-filled band Mott-Hubbard insulator of
Fig.~4(a), reached after Jahn-Teller distortion
\cite{Iwasa03a,Capone09a}.

Within a dynamic mean field theory (DMFT) of the pressure-induced
antiferromagnetism-to-SC in Cs$_3$C$_{60}$ \cite{Capone09a}, pairing
arises from the combined effects of Hubbard $U$ and Jahn-Teller
interaction within the effective $\frac{1}{2}$-filled band of
Fig.~4(a), at the interface of an insulating and a metallic phase.  We
recall that the DMFT approach to SC in the CTS within the effective
$\frac{1}{2}$-filled band Hubbard model was very similar. It is now
known that SC in this latter case was an artifact of the mean-field
approximation \cite{Clay08a}. We believe that a similar criticism
applies also to the existing theory of SC in the fullerides.

{\it The difficulty in arriving at a theory of SC arises from the bias
  that the orbital occupancies are the same in the antiferromagnetic
  and superconducting phases.} We suggest that pressure induces a new
orbital ordering, with C$_{60}^{3-}$ configurations as shown in
Fig.~4(b), where two degenerate $\frac{1}{4}$-filled bands bands are
obtained. This orbital ``reordering'' will be driven by the lower
total energy of two $\frac{1}{4}$-filled bands compared to that of the
single effective $\frac{1}{2}$-filled band, for larger bandwidth. Each
of the $\frac{1}{4}$-filled bands can now form its own paired-electron
liquid.

\section{Conclusion.} 
Our goal here was to show that the correlated $\frac{1}{4}$-filled
band offers a common description of seemingly very different
classes of materials as well as a plausible mechanism of SC. The key
step is the realization that antiferromagnetism in the 
semiconducting state does not necessarily mean that the electronic
structure of the superconductor is derived from the same
configuration. Bandwidth-driven transition to a
different state that is more appropriately described as $\frac{1}{4}$-filled 
can occur in the superconducting state \cite{Clay11a}. This has been explicitly shown in the context of the CTS. 
Work is in progress to demonstrate the same within the models of Figs. 3 and 4.
The proposed mechanism offers, (a) a way to
arrive at a single unified theory of SC for the CTS, with different kinds of proximate insulating
states, and (b) understanding of the strong role of electron-phonon
interactions, and yet large T$_c$/T$_F$, in
the spinels and the fullerides. In both spinels and fullerides, dynamic
Jahn-Teller phonons will play a strong role in the correlated
superconductor. However, this is not a signature of BCS pairing.

\begin{acknowledgement}
This work was partially supported by  DOE Grant No. DE-FG02-06ER46315.
\end{acknowledgement}

\end{document}